# AI in Telemedicine: An Appraisal on Deep Learning-based Approaches to Virtual Diagnostic Solutions (VDS)

Ozioma Collins Oguine and Kanyifeechukwu Jane Oguine

Department of Computer Science, University of Abuja, Nigeria

## ABSTRACT

*Advancements in Telemedicine as an approach to healthcare delivery have heralded a new dawn in modern Medicine. Its fast-paced development in our contemporary society is credence to the advances in Artificial Intelligence and Information Technology. This paper carries out a descriptive study to broadly explore AI's implementations in healthcare delivery with a more holistic view of the usability of various Telemedical Innovations in enhancing Virtual Diagnostic Solutions (VDS). This research further explores notable developments in Deep Learning model optimizations for Virtual Diagnostic Solutions. A further research review on the prospects of Virtual Diagnostic Solutions (VDS) and foreseeable challenges was also highlighted. Conclusively, this research gives a general overview of Artificial Intelligence in Telemedicine with a central focus on Deep Learning-based approaches to Virtual Diagnostic Solutions.*

## KEYWORDS

*Biomedical imaging, Telemedicine, Smart Healthcare, Medical Imaging, AI, Virtual Diagnostic Solutions.*

## 1. INTRODUCTION

Healthcare and Medicine are areas of modern society which has gained quite an outstanding level of research attention given current antecedents of virus outbreaks and spikes in anomalies regarding human health. Over the years, advancement in Artificial Intelligence and its resonating research areas such as Telecommunication and information technology has stirred up questions and advanced solutions regarding Human health. Affirmatively, we can infer that these improvements have notably impacted the medical and healthcare delivery scale and quality. However, healthcare access and delivery have struggled extensively to meet anticipated simultaneous prospects, as is the situation in many parts of the world, predominantly in underdeveloped and developing nations. A significant reason for this decline is the ever-increasing number of healthcare users and medical patients leading to the overutilization of medical resources such as healthcare providers, medical staff, and access to medical infrastructures. Hence, a more suitable and sustainable healthcare delivery approach was necessary to eliminate the issues arising from overpopulation and access to healthcare infrastructures.

A notable convergence of Machine Learning, Robotics, Telecom, computational neuroscience, and cloud computing has established a new infrastructure for global healthcare delivery known as **Telemedicine**. According to Khemapech et al., "Telemedicine is the delivery of health care services, with significant consideration of distance in service delivery and accessibility by all stakeholders as key variables, using information and communications technologies" [1]. An inferred purpose of this innovation is to exchange valid information for diagnosing, treating, and preventing disease and injuries. Also, for research and evaluation, continuing education of health care providers and users, all aimed at advancing the healthcare systems of individuals and their communities. Although research on Telemedicine has been ongoing for decades, the emergence of the COVID-19 pandemic has reinvented its usage instantaneously, resulting in its scalability in fully implementing essential health and safety protocols in service delivery. This research area is an innovation that transcended from being a convenient alternative for technically savvy patients to the mainstay for healthcare delivery today across nations of the world, a reality that may continue long after the COVID pandemic. Telemedicine is the most suitable and sustainable approach to delivering healthcare services to patients by incorporating technological advancements in affordable and low-cost implementations. It also eliminates factors that adversely affect privileges to healthcare by strategically expanding virtual solutions to accommodate the growing populace and integrating smart tech to facilitate healthcare access and efficiency. The telemedicine framework has enabled collective improvements in Virtual Diagnostic Solutions, as shown in Fig 1.

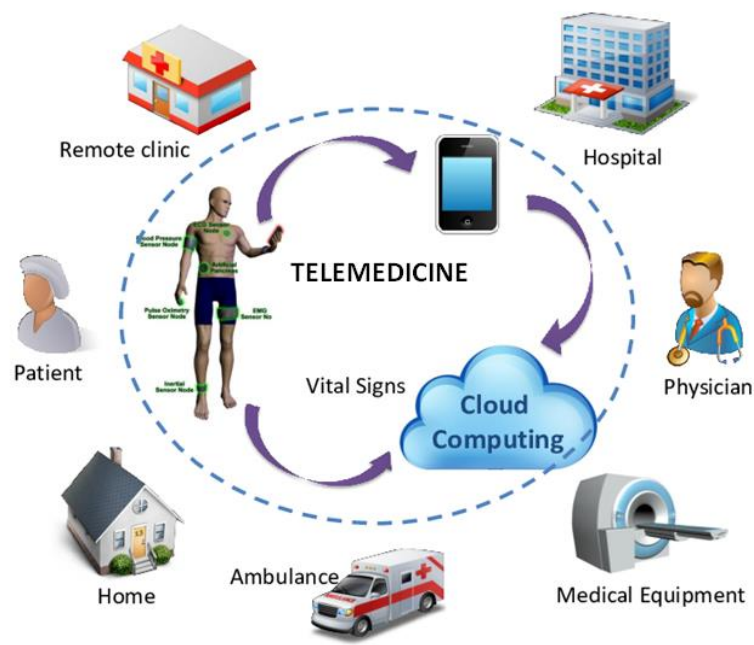

Fig 1. Telemedicine Infrastructure and Components

**Deep Learning:** Deep learning can be referred to as a modern state-of-the-art mechanism for computational processes applied to solving contemporary problems. Its potential for drawing patterns and insights from a massive amount of data boasts its wide adoption.

**Virtual Diagnostic Solutions (VDS):** This is a systematic approach to undertaking medical prognosis and administering or proffering medical solutions with little or no supervision from medical Professionals.

## 2. STAKEHOLDERS IN TELEMEDICINE

Telemedicine as an infrastructure for healthcare improvement entails the participation of four significant stakeholders, namely:

**Patients (Key Stakeholders):** These are the receivers or users of healthcare services. They are perhaps the most critical stakeholder to consider when thinking of approaches to Telemedicine.

**Medical Professionals:** These are the next most essential stakeholders in Telemedicine. Their knowledge area and research lay the foundation for building a telemedical Infrastructure.

**Developers/ Tech Experts:** These stakeholders are necessary because they constantly research innovative ways to build solutions to medical problems (Middlemen between Patients and Medical Professionals).

**Policy Makers:** These are folks who formulate and regulate laws that guide the implementation and utilization of Telemedicine

Telemedicine is currently gaining legislative and regulatory support in most developed countries; there have been no nationally representative estimates on its implementation by physicians and health practitioners across all medical specialties [2]. "To tackle this information gap, the American Medical Association (AMA) conducted a study in 2016 that surveyed 3,500 physicians who provided needed data to help assess potential barriers and create strategies to promote telemedicine adoption. The data analysis report from the AMA's 2016 Physician Practice Benchmark Survey clearly described the Physician-to-Patient ratio of Virtual Diagnostic Solutions (VDS) in Telemedicine, as shown in Fig. 2. Consequently, this paradigm also proved efficient in maintaining and sharing patients' medical records among hospitals and medical institutions.

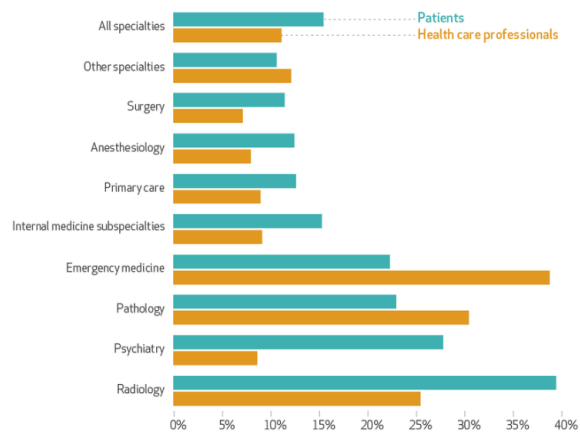

Fig 2. AMA 2016 Data Analysis on Patient to Healthcare Professional ratio in Telemedicine

The advancement of Artificial Intelligence and its research areas, such as the Internet of Things (IoT), Machine Learning, Image processing, and Deep Learning, has highly improved the level and quality of services provided by Telemedical software and applications. Researchers have and are currently up scaling on services and innovative solutions such as medical prognosis, Analytical Medicine, Robotic Surgery, DNA (genome) Sequence Analysis (DSA), Drug research and discovery, Medical Data Security, Clinical Trials, Medical Risk Prediction, emergency services, and Medical Image Analysis (brain monitoring, computer tomography, and radiology) offered by this remarkable means of healthcare delivery. Amid all the solution-oriented approaches put forward to dress health challenges, medical imaging is one area that has seen more promising results and prospects. X-Ray, Ultrasound (US), MRI, and CT Scanners have been interfaced with computers to transfer medical images to the remote center for careful analysis and early detection of medical abnormalities [3].

This paper explicitly discusses the implementation of artificial intelligence in Telemedicine with a more holistic view of Annotation-Efficient deep learning models for Medical Imaging in Virtual Diagnostic Solutions (VDS). It also elaborates on how data (input images) from medical imaging equipment are passed to sophisticated algorithms for accurate Diagnosis and treatment of Imaging-related ailments. Furthermore, application scenarios of medical Imaging models for VDS are subsequently highlighted and reviewed to understand its challenges and perceived prospects.

## 3. ARTIFICIAL INTELLIGENCE IN TELEMEDICINE

Artificial intelligence is a broad research field that facilitates machine simulation of human intelligence and behaviors ranging from learning to problem-solving. Its rapid growth and development in the past decade are due to its vast implementation in all human endeavors. Given the recent upsurge in big data generation, powerful computing coupled with refined computational models and algorithms, developments in AI have accelerated exponentially [4]. This trend has flagged the emergence of subfields such as Machine Learning (ML), Natural Language Processing (NLP), AI voice technology, Medical Imaging, AI assistants, Computer Vision, and robotics.

Telemedicine is a resurging innovation that has gained wide adoption in most developed and developing countries to Fastrack the accessibility to sustainable healthcare. It has achieved quite a laudable level of attention and research over the last two decades as ongoing studies are exploring ways to improve the existing infrastructure to a State-of-the-Art (SOTA). For efficient healthcare provision in contemporary societies, humans and machines have complimented each other in effectively delivering healthcare services through Virtual Diagnostic Solutions (VDS), thus providing a platform that bridges the gap between communication and accessibility to medical services. AI in Telemedicine has seen rapid adoption primarily based on the volume of medical data (Big data) recurrently generated. According to Ozioma et al., traditional data-handling techniques have proved ineffective in utilizing these data to obtain viable medical insights or solutions, given the gargantuan nature of the data [5]. Hence, sophisticated demonstrations of Artificial Intelligence approaches and models in Medical Diagnostics are becoming relatively popular. The adaptability and flexibility of these AI-based approaches and models have also driven the necessity for their implementation.

Andressa et al., amongst other scholars (see table 1), proposed an architecture that relies on fingerprinting and FLIPER framework to Fastrack the versatility and interconnectivity of healthcare applications. They also anticipated impact of their research was to enable quick, customizable resources that meet the level of reliability required for AI in Smart-health applications [6].

Table 1. Comparative Table describing Researches implementation of AI in Telemedicine [7]

| Research Article | Year | Trend Category | Methodology |
|---|---|---|---|
| A Predictive Model for Assistive Technology Adoption for People with Dementia [8] | 2014 | Information Analysis and Collaboration | KNN and other data mining algorithms were utilized to analyze the behavior of patients with dementia and their adaptation to technology. |
| A Telerehabilitation Application with Pre-defined Consultation Classes [9] | 2014 | Healthcare Information Technology | Issues of Telemedicine under low-bandwidth network conditions were addressed using customized consultation classes demonstrating rehabilitation practices with preset parameters and a bandwidth adaption algorithm |
| An application of fuzzy systems was used to identify the best course of action for a given situation to Monitor Mobile Patients by Combining Clinical Observations with Data from Wearable Sensors [10] | 2014 | Intelligent Assistance Diagnosis | "This paper explores principled machine learning approaches to interpreting large quantities of continuously acquired, multivariate physiological data. Early warning of serious physiological determination was done using wearable patient monitors, such that a degree of predictive care may be provided." |
| Ankle Rehabilitation System with Feedback from a Smartphone Wireless Gyroscope Platform and Machine Learning Classification [11] | 2015 | Patient Monitoring | This study uses a smartphone application, a wireless gyroscope platform, machine learning, and 3D printing to record usage and effects of therapy on an ankle and measure the strategy's efficacy. |
| Intelligent decision systems in Medicine -a short survey on medical Diagnosis and patient management [12] | 2015 | Intelligent Assistance Diagnosis | The study presents "a short review of some current Machine Learning algorithms (neural networks, genetic algorithms, support vector machines, Bayesian decision, k-nearest neighbor, etc.) used for automated Diagnosis of different major diseases, such as breast, pancreatic, and lung cancer, heart attacks, Diabetes." |
| Smartphone-Based Recognition of States and Changes in Bipolar Disorder Patients [13] | 2015 | Intelligent Assistance Diagnosis | This paper proposes a system of using a smartphone-sensing wearable device to evaluate the behavior and recognize depressive and manic states of patients with bipolar disorder. |
| An Effective Telemedicine Security Using Wavelet-Based Watermarking [14] | 2016 | Information Technology | This paper proposes an algorithm that embeds and reads digital wavelet watermarks on medical images to secure confidentiality. |
| Mobile Cyber-Physical Systems for Health Care: Functions, Ambient Ontology and e-Diagnostics [15] | 2016 | Patient Monitoring | This paper proposes the use of a monitoring system embedded in wearable devices for the doctor or family members to receive updates on the patient's status. |

| Detection of Fetal Electrocardiogram through OFDM, Neuro-Fuzzy logic, and Wavelets Systems for Telemetry [16] | 2016 | Patient Monitoring | This study uses a neuro-fuzzy logic system to monitor and detect the exact electrocardiogram and other signals of a fetus inside an abdomen. |
|---|---|---|---|
| Using CART for Advanced Prediction of Asthma Attacks Based on Telemonitoring Data [17] | 2016 | Intelligent Assistance Diagnosis | This study created an algorithm with data from a home-based telemonitoring system to predict asthma exacerbation. |
| A Wireless Continuous Patient Monitoring System for Dengue: Wi-Mon [18] | 2017 | Patient Monitoring | The paper presents "a wireless monitoring system for patients who need continuous monitoring, using the Wireless Body Area Network (WBAN) concept." |

This research also aims to describe and evaluate the impact of Artificial Intelligence in Telemedicine. Some documented implementations of this powerful paradigm are discussed below:

**Medical Imaging:** This generally entails training AI models with images of medical scans that have been scientifically collected and stored in data repositories and databases. AI has significantly reduced the cost and time involved in analyzing scans through advanced deep learning models to diagnose health disorders. Hence, potentially allowing more scans to be taken and improving prognosis proficiency and accuracy [19].A vast amount of resources and research has been expended in developing this field, as will be accentuated later in this paper. These researches have paved the way for State-of-the-art detection methodologies of medical ailments such as Brain tumors, skin and breast cancer, Pneumonia, and eye diseases.

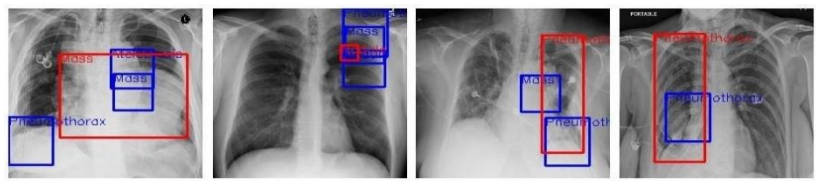

Fig 3. Covid-19 Detection in Lungs using Multibox SSD Model [20]

**Echocardiography:** Heartbeat patterns and coronary heart disease diagnosis and detection utilizethe Ultromics system, an AI framework trialed at John Radcliffe Hospital in Oxford, to analyze echocardiography scans [21].

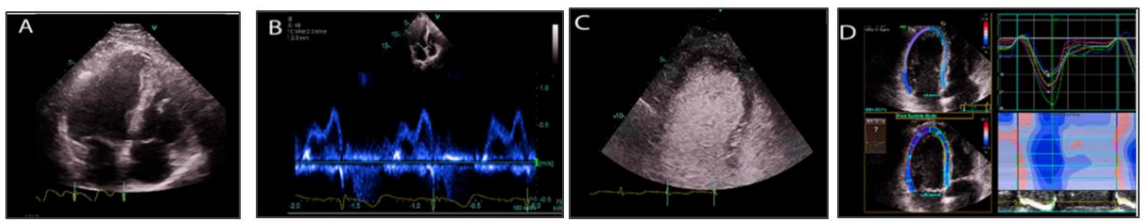

Fig 4. Classification of Echocardiograms using Deep Learning [22]

**Screening for Neurological Conditions:** Several AI models are being developed and employed in speech patterns analysis to predict psychotic episodes, Schizophrenia, recognize and monitor symptoms of neurological disorders such as Parkinson's disease [23].

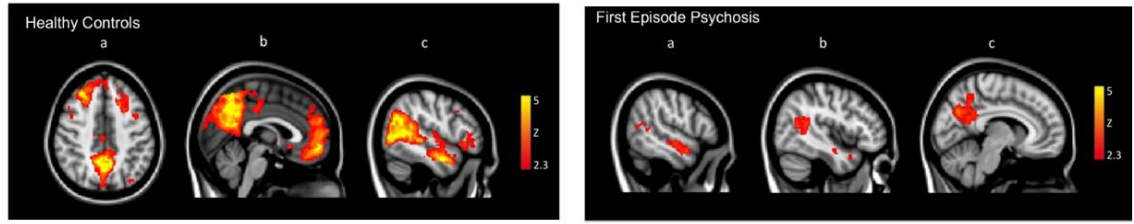

Fig 5. fMRI Study in Individuals with First-Episode Psychosis [24]

**Emotion Recognition for Psychological Prognosis:** Artificial Intelligence Deep learning models have proven efficient in recent times in observing and predicting emotions through individual reactions and psychological states at a given time. Ozioma et al. opined in their research on Facial Expression Recognition that Emotions are fundamental in human communication, driven by the erratic nature of the human mind and the perception of relayed information from the environment [25]. They proposed a hybrid deep learning model that makes real-time predictions of a person's emotional state, categorizing the individual's emotion into one of seven classes. The need for novelties in advancing this research field stems from the alarming rate of Emotion disorders, Suicide, Post-traumatic stress disorders (PTSD), and Psychological/Mental breakdown suffered by people in our contemporary society.

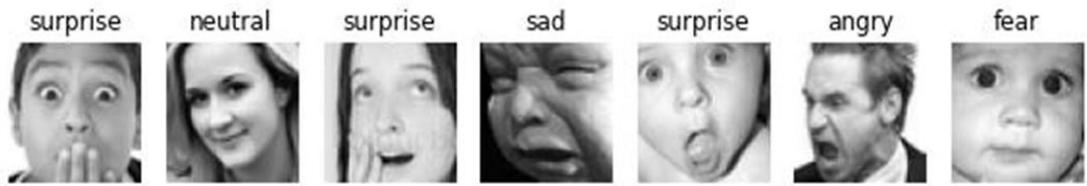

Fig 6. Facial Emotion Recognition using Hybrid Deep Learning Model [25]

**Teleradiology:** "Telecommunication is employed to transmit digital radiological images, like X-rays, Computed Tomograms (CTs), and Magnetic Resonance Images (MRI's) across geographical locations for interpretation and consultation" [26]. Reducing financial costs is one of the primary benefits of Teleradiology as it significantly reduces constraints in accessing radiological images, reports, and feedback between health professionals and patients.

Teleradiology is a crucial means for optimizing radiology workflow by sending the images to the radiologist rather than traditionally going to the radiology facility. By its very nature, Teleradiology is an efficient and high-quality manner by which patients' images can be interpreted and diagnosed by qualified specialists. Cloud services are primarily employed in this Telemedical service, where health stakeholders can utilize numerous privileges from the cloud, thus, upscaling the quality of radiology services. AI in Teleradiology will also enable the sharing of clinical information, medical imaging studies, and patient diagnostics [27] between patients and healthcare professionals.

**Teledermatology:** This service implements Telecommunication to transmit medical information concerning skin conditions (e.g., tumors) for interpretation and consultation. According to Eedy and Wotton, "Teledermatology model has received extensive advocate as a healthcare delivery model that may reduce inequalities encountered in the utilization of overstretched dermatological services, with implementation concentration ranging from remote to isolated communities" [28]. Landow et al., in their research, highlighted four factors that have stirred a relative increase in face-to-face appointments which teledermatology strategies have tackled: (1) effective pre-selection of patients for teleconsultation, (2) high-quality photographic images, (3) dermoscopy if pigmented lesions are evaluated, and (4) adequate infrastructure and culture in place to implement teleconsultation recommendations [29].

## 4. APPLICATIONS OF DEEP LEARNING-BASED DIAGNOSTIC SOLUTIONS IN TELEMEDICINE

State-of-the-art Deep Learning models have seen advancements in methodologies over various medical problems such as object detection, recognition and segmentation in computer vision, voice recognition, and genotype/phenotype prediction. Telemedicine employs deep learning models in several Virtual Diagnostic Solutions today. While early studies focused on 2D medical images, such as chest X-rays, mammograms, and histopathological images, recent studies are looking toward applying sophisticated deep learning models to volumetric medical images.

CNNs form the basis for most State-of-the-art medical imaging DL models, which have gained wide prominence since achieving impressive results at the ImageNet [30] competition in 2012. Akkus et al. opined that CNNs remain a popular choice of DL approaches to image processing given their laudable tendency to weight sharing across convolutional layers or feature maps, in contrast, to fully connected ANNs [22]. And a rational reason for this was that, for 2D/3D image processing, ANNs utilize heavy computational processing that consumes a relatively high amount of GPU memory.

Several scholars have conducted evaluation studies and proposed several methodologies for different Virtual Diagnostic Solutions employing medical Imaging to solve health-related issues ranging from collecting image data to evaluating and diagnosing medical ailments. Qin et al. utilized Computer-aided Detection (CAD) in chest radiography based on contrast enhancement and segmentation in diagnosing various lung diseases such as early lung cancer, Pneumonia, Tuberculosis, and, more recently, lung inflammation levels caused by Covid-19 [31]. Numerous well-known DCNN architectures tested by Shin et al. emphasized the efficiency of transfer learning approaches in CT patch-based thoracoabdominal lymph node detection and ILD classification [32]. A Recurrent full Convolutional Neural Network (RFCNN) proposed by Poudel et al. was used to segment the left ventricle from cardiac MR images [33]. Hosseini-Asl et al. proposed a 3D CNN model to determine the progression of Alzheimer's disease from structural brain MR images [34]. In the method, they employed a transfer learning approach that utilized pre-trained weights of features from a Computer-Aided Engineering (CAE) with a small number of source domain images to fine-tune the target domain data to train the actual classification model. Yu et al. proposed a 3D volumetric CNN for prostate segmentation on MR images through U-net expansion, initially used for 2D biomedical image segmentation. They also added residual connections to combine multiple-scale information [35, 36]. An alternative study proposed a slice-level classification model to detect Interstitial Lung Diseases (ILD) from chest CT scans [37]. Jamaludin et al. detected several diseases simultaneously from spinal MR images through a trained multi-task learning model. They visualized salient regions in the image for corresponding predictions as 'evidence hotspots' as seen in Fig.7 [38].

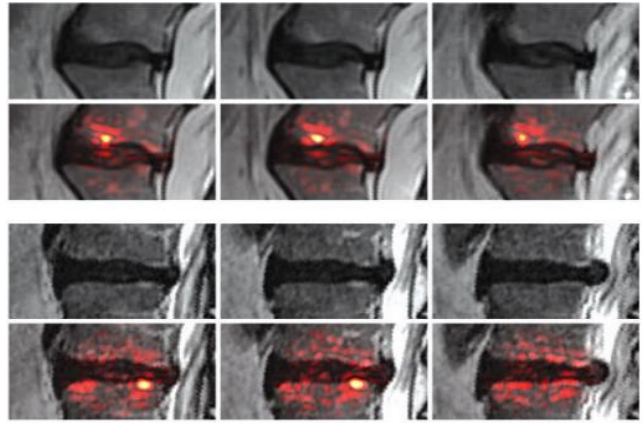

Fig 7. "Evidence hotspot" Visualization in Spinal MRI [38]

## 5. PROSPECTS OF AI-BASED VIRTUAL DIAGNOSTIC SOLUTIONS (VDS) IN TELEMEDICINE

Early Diagnosis and treatment of diseases have been a primary focus of modern-day healthcare infrastructure. In previous sections above, this research has significantly highlighted some outstanding advancements and implementation of the Deep Learning approach of Artificial Intelligence to improve Virtual Diagnostic Solutions. This review paper also seeks to forecast based on the current state-of-the-art prospects of this frequently evolving paradigm. A crucial benefit of this paper is to buttress the promises of AI in future healthcare development.

**Early Diagnosis and Treatment:** Diagnosis and treatment have ensured the advancement of research conducted in the field of Medicine today. With the current pace in development, a significant prospect of AI in Telemedicine is seen to be geared towards establishing and improving early diagnosis and treatment mechanisms of numerous medical conditions. Jacobsmeyer showed the effectiveness of AI in the early detection and management of infectious diseases and epidemics such as Water contamination, worldwide virus outbreaks, etc. [39]. Research Labs and tech firms are currently pushing the limits on a proactive scale in ensuring this, as not a week goes by without the introduction of new approaches to medical solutions using AI or Big data with improved accuracy and precision.

AI systems will also become more advanced in engaging in broader medical problem-solving tasks with little human supervision or control. Hence, advanced ethical learning as imposed by state-of-the-art deep learning models will play a crucial role in the effective implementation strategy and adoption of Artificial Intelligence approaches in Virtual Diagnostic Solutions.

**Independence:** A major gap identified in traditional Medicine is the inability of patients to access healthcare when, where, and how they want it. Deep learning approaches to Telemedicine have gone far beyond filling this gap by empowering patients to systematically, thematically, and objectively evaluate symptoms and proffering possible solutions through virtual diagnostic solutions. Another significant advantage is the noteworthy reduction in the strain on medical resources and healthcare professionals. While concerns have been raised regarding the negative impact of AI solutions on Telemedicine [19], several notable research has been observed to acknowledge its importance and necessity in delivering cost-effective, high quality and accessible healthcare services [40, 41].

**Improved Clinical and Therapeutic Coordination:** Not only have the efficiency and advancements of AI ensured quality in the level of medical potential, but they have also shown a remarkable promise in the level of service delivery mechanisms. As more implementations of Virtual Diagnostic Solutions are utilized, huge amounts of data are also generated from healthcare professionals and patients. Over time, this aggregate amount of data will create an abundance of information on stakeholders involved in Telemedicine utilization. A foreseeable impact of this will be the development of systematic and coordinated clinical therapeutic processes and services that ensure that VDS prognosis are subjected to second opinion evaluations to establish accurate Diagnosis, the improved continuity in medical care through research and training mechanisms, simplification of medical procedures and the analysis of medical inconsistencies and patterns in patient health to foster the development of new patient-centered models.

**Scalability:** Global healthcare witnessed new dawn with the discovery of Telemedicine which ensured that medical solutions could be provided outside hospital buildings. Virtual workflows and technologies have been set up over time to create optimal infrastructures to implement this process. The scalability of Virtual Diagnostics Solutions used in Telemedicine has been a significant metric of development in our contemporary society. Evolutionary standards in AI have ultimately ensured the continuous upgrade of Telecommunications, software, and digital technologies used to implement virtual and real-time diagnostic solutions. A relative expansion in the number of telemedical services provided by Virtual Diagnostic Solutions has spiraled the frequent adoption of this paradigm in tackling global healthcare challenges such as epidemics, pandemics, and critical health conditions.

Another reason for improvements in scalability is the ever-changing needs of healthcare stakeholders to meet state-of-the-art next-generation service level requirements. Scalability could entail efficient resource allocation, enhanced data integrated hospital-grade wearable devices, expansion of patient management software systems, robust databases, cloud services, and data-sharing infrastructures. As we advance, a perceived rise in the adoption of the telemedical model of healthcare delivery will no doubt facilitate accessibility and efficiency with the sole purpose of bringing a paradigm shift to Global health solutions.

## 6. CHALLENGES OF AI-BASED VIRTUAL DIAGNOSTIC SOLUTIONS (VDS) IN TELEMEDICINE

Despite the numerous benefits promised by the integration of AI in Virtual Diagnostics Solutions (VDS) to provide geographically accessible, affordable, acceptable, and quality healthcare, there have also been challenges mitigating its full-scale adoption and development potential. As stated earlier, this paper will highlight some key barriers facing the advancement of Virtual Diagnostic Solutions in Telemedicine.

**High-Cost:** Although Virtual Diagnostic Solutions have been leveraged to reduce the high cost of healthcare accessibility, this is only valid from a patient's perspective. While this issue might seem improbable in our contemporary society, it is pertinent to note that '*Cost*' sums up the general resources required to develop, implement, maintain and advance Virtual Diagnostic Solutions in Telemedicine. These resources include but are not limited to human, technical, financial, and academic resources that have become relatively expensive to acquire and employ over time. Cost implications for efficient initiation and delivery of telemedical research and projects are seen as economic excess and, as such, given less consideration. This effect has led to the relatively reduced advocacy for adopting telemedical solutions in most developing and underdeveloped countries. From an AI perspective, huge amounts of technical resources, skillsets, and training are required to build robust Telemedical deep learning models for Virtual Diagnostic Solutions. Financial resources are expended on the acquisition, installation, utilization, and maintenance of telemedical equipment. On a more holistic scale, this factor is the most critical, responsible for the decline and a foreseeable reduction in VDS development and adoption in Telemedicine.

**Unavailability and Underdevelopment of Technical Infrastructures:** Technology, Telecommunication, and Artificial Intelligence advancements are the backbone for functioning state-of-the-art Telemedical solutions, as have been elucidated in earlier sections of this paper. Hence, the robust nature of Telemedicine has necessitated the requirement of sophisticated infrastructures to ensure efficient and effective development and deployment. For instance, Tele-ophthalmology, Teleradiography, and real-time emergency consultation, amongst others, are some of the telemedical services which require heavy computing technologies and fast internet connectivity. Systematically excellent and diverse medical data are also needed for training Machine Learning (ML) and deep learning models to enhance effective generalization. Technology literacy levels also play a part in ensuring telemedical services' smooth implementation and sustainability by abstracting the process workflow to patients and medical professionals. The unavailability of these requirements is predicted to pose a significant gap in harnessing the potential of Telemedicine. On a more holistic scale, a lack of technical infrastructures will more likely hinder the development of Telemedicine, thereby causing a decline in the progress achieved so far. Despite the encouragement by the WHO encouraging the adoption of Telemedical innovations by member states, a major drawback emanating from this challenge has been observed in underdeveloped and developing countries.

**Reliability Issues:** A significant concern regarding the adoption and implementation of Virtual Diagnostics Solution is its reliability and generalization ability. Given this issue's validity, several Virtual Diagnostics solutions have come under severe scrutiny, raising questions and sentiments concerning their utilization. Reliability is a dominant parameter in determining the rate of adoption and research in Telemedical solutions. Hence, at any slightest detection of untrustworthiness or inconsistency, Telemedicine could lose the attention and participation of key stakeholders. Error-prone VDS solutions have stereotypically discouraged total reliability in the potentials of Telemedicine. While a significant level of this challenge is due to lapses in technicalities and operational models, several other factors can also contribute to this issue, such as stakeholders' resistance to accepting change, Digital illiteracy leading to poor awareness of modern tech, cultural perceptions, and malpractice liabilities.

**Ethical Violations, Confidentiality, and Privacy issues:** Ever since the revolutionization of the Internet, privacy and ethical policy stability have been a critical challenge among internet users. Just like Medical professionals, patients (VDS users) need orientation and training on data access, privacy and protection measures when utilizing services on Virtual Diagnostic Solutions. For efficiency of Telemedicine, data privacy such as Doctor-Patient confidentiality, training, and licensing of personnel is expected to be strictly adhered to. However, stability in providing robust security infrastructures and interoperability features to tackle these challenges, coupled with the ever-rising trend of cyber-crimes, has hindered the trustworthy adoption of Telemedicine as an effective approach to healthcare access. Another propelling reason for this challenge is the result of insufficient legal policies, guidelines, and Standard Operating Procedures (SOPs). In addition to the absence of defined policies and regulatory procedures, a lack of international regulatory uniformity has stirred several controversies regarding Telemedical services and solutions. While AI's propensity for good has been established in earlier sections of this paper, Malpractice liability is another factor to consider. This issue has necessitated holistic reflections on AI's dual potentials by governments, Researchers, and Engineers developing Telemedical solutions.

**Continuity/Sustainability:** The recent global pandemic has necessitated the sustainability of advanced medical technologies. As more medical issues arise, so should the pervasiveness of healthcare approaches employed for treatment purposes. A major drawback, especially in underdeveloped and developing countries, is their incapability to adequately sustain and encourage advancements in Telemedicine either through research or system analysis. Several factors can be attributed to this enthusiastic acceptance of the already existing Virtual Diagnostic Solutions, formulation of ethical and privacy policies, standardization of technological equipment, skill set and infrastructures, cost-effectiveness and coordination, etc. In light of this, the sustainability and continuity of this ever-growing trend hinge on the improvements of all participating stakeholders in creating stable infrastructures and approaches to tackle existing challenges and foster the growth and application of AI in Virtual Diagnostic Solutions (VDS).

## 7. CONCLUSION

Artificial Intelligence has no small impact on Global health in our contemporary society. In most developed and some developing countries, Telemedicine has gained popularity for its benefits in improving healthcare access, reducing healthcare costs, and enhancing the quality of healthcare services. These benefits necessitated the rapid paradigm shift from the usual traditional healthcare (provider-centric) infrastructures to more robust (patient-centric) methodologies and infrastructures, given the tremendous pressure on healthcare providers to provide affordable, accessible, and quality healthcare services. This paper holistically discussed the growth and revolution of Telemedicine as a modern research field; it also introduced a broad insight towards implementing Artificial Intelligence in Telemedicine with specific inclinations toward services provided by Virtual Diagnostic Solutions (VDS). A review of works of literature from researchers citing several Deep Learning Approaches employed in detecting and treating several medical ailments was also discussed in this paper, recommending the importance of deep learning in the advancement of AI services. Several Applications of Artificial Intelligence in sustainable healthcare provision and access were also reviewed and described.

The overall significance of this paper is to throw more light on the importance of ***DEEP LEARNING-BASED*** AI methodologies in advancing state-of-the-art Virtual Diagnostic Solutions in Telemedicine. Consequently, this paper enumerated and discussed crucial prospects and challenges associated with incorporating, implementing, adopting, and advancing Artificial Intelligence in Telemedicine. Further holistic research on advanced Telemedical approaches is encouraged in tandem with these views.

## ABBREVIATIONS

IoT - Internet of Things  
DSA - DNA Sequence Analysis  
VDS - Virtual Diagnostic Solutions  
ML - Machine Learning  
CT - Computed Tomography  
NLP - Natural Language Processing  
CAE - Computer-Aided Engineering Network  
ILD - Interstitial Lung Diseases  

CAD - Computer-Aided Detection  
AI - Artificial Intelligence  
MRI - Magnetic Resonance Imaging  
US - Ultrasound  
AMA - American Medical Association  
SOTA - State-of-the-Art  
RFCNN - Recurrent Fully Convolutional Neural Network  
SOPs - Standard Operating Procedures  

## DISCLOSURE

The authors declare that they have no competing interests with anyone in publishing this paper.

**AUTHOR CONTRIBUTIONS**

All authors made substantial contributions in conscripting the paper and revising it critically for important intellectual content; agreed to submit it to the current journal; and final approval of the version.

**AUTHORS**

**Ozioma Collins Oguine** is a Graduate Research Assistant at the University of Abuja, Nigeria. He graduated from the same University with First Class Honors (Summa cum Laude), top 1% from the Department of Computer Science. His research interests are Machine/Deep Learning, Computer Vision, Robotics, and Human-Computer Interaction (HCI). He is a Member of the Intelligent Automation Network (IAN), Black in AI, Black in Robotics, an illustrious member of the International Society of Engineers (IAENG) in Artificial Intelligence and Computer science, and an Associate Member of the British Computing Society (BCS).

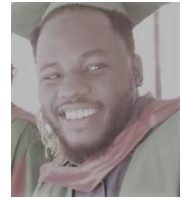

**Kanyifeechukwu Jane Oguine** is a Graduate Research Assistant at the University of Abuja, Nigeria. She graduated from the same University with First Class Honors (Summa cum Laude), top 2% from the Department of Computer Science. Her research interests are Machine/Deep Learning, Computer Vision, Computational Algorithm, and Human-Computer Interaction (HCI). She is a Member of the Intelligent Automation Network (IAN), Black in Robotics, also a notable member of the International Society of Engineers (IAENG) in Artificial Intelligence and Computer science, and an Associate Member of the British Computing Society (BCS).

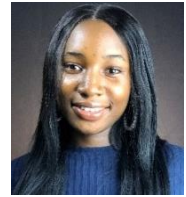